\documentclass[submission,copyright,creativecommons]{eptcs}
\providecommand{\event}{WCSI 2010}

\usepackage{graphicx} 
\usepackage{breakurl}

\title{A Reusable Component for Communication and Data Synchronization in Mobile Distributed Interactive Applications}

\author{Abdul Malik Khan and Sophie Chabridon
\institute{Institut T\'el\'ecom, T\'el\'ecom Sudparis, CNRS UMR SAMOVAR\\
9 Rue Charles Fourier, F-91011 \'Evry, France\\
}
\and
Antoine Beugnard
\institute{Institut T\'el\'ecom, T\'el\'ecom Bretagne\\
Computer Science Department, F-29238 Brest cedex 3, France}
\email{\{Abdul\_Malik.Khan,Sophie.Chabridon,Antoine.Beugnard\}@institut-telecom.fr}
}
 
\def\titlerunning{A Reusable Component for Data Communication and Synchronization}
\def\authorrunning{A.M. Khan, S. Chabridon \& A. Beugnard}
 
\begin{document}
\maketitle

\begin{abstract}
In Distributed Interactive Applications (DIA) such as multiplayer games, where many 
participants are involved in a same game session and communicate through a network, they may have an 
inconsistent view of the virtual world because of the communication delays 
across the network. This issue becomes even more challenging when communicating 
through a cellular network while executing the DIA client on a 
mobile terminal. Consistency maintenance algorithms may be used to obtain a 
uniform view of the virtual world. These algorithms are very complex and hard 
to program and therefore, the implementation and the future evolution of the 
application logic code become difficult. To solve this problem, we propose 
an approach where the consistency concerns are handled separately by a 
distributed component called a synchronization Medium, which is responsible for 
the communication management as well as the consistency maintenance. 
We present the detailed architecture of the synchronization Medium
and the generic interfaces it offers to DIA.
We evaluate our approach both qualitatively and quantitatively. 
We first demonstrate that the synchronization Medium is a reusable component 
through the development of 
two game applications, a car racing game and a space war game.
A performance evaluation then shows that the overhead introduced by the 
synchronization Medium remains acceptable.\\


 \end{abstract}

\section{Introduction}
In multiplayer games, where many players take part in playing a game across a network, the communication of information is necessary between the different players to keep them updated about the global state of the game. This information is passed from one player to another via a network with a certain non-zero delay. Because of this network latency, information takes time to reach one player from another one. Also, in Distributed Interactive Applications (DIA) \cite{DIS} such as multiplayer games, the state of the game changes not only because of user actions which occur at discrete time instants, but also with the passage of time. DIA are thus characterized as continuous applications and require specific synchronization techniques \cite{Mauve}. When the information is being passed from one player to another one across a network, the game state of the sender might have changed before the information reaches its destination and hence this information becomes obsolete. Therefore, consistency management algorithms are necessary to reach a 
common state at each client and to hide the network latency to the players. The problem of the network latency becomes even more challenging when the players are connected via a wireless network, where the delay can be of many seconds. 
Also, there are more chances of jitters and messages losses in wireless networks. So to achieve a consistent state acceptable to the application users, complex consistency control mechanisms are required. Usually these mechanisms and the related algorithms are developed by the game programmer and are embedded in the game logic. This makes difficult the development of the game and its evolution. 

Based upon the concept of Medium \cite{eric}, we have proposed in {\cite{ours07} to separate these algorithms from the game logic and embed them into a middleware layer called Synchronization Medium. This Synchronization Medium is responsible for the consistency management and the synchronization of the game information and also for hiding the network latency. Figure \ref{fig:Medium-layers} shows a Synchronization Medium used by a game application. The Medium is a component that is installed on every client and lies between the game application and the underlying network infrastructure. In the figure, the set of all the PlayerManagers on all the clients constitutes a Synchronization Medium. A Synchronization Medium offers services to the game client and requires services provided by the game application. The Synchronization Medium can be seen as an abstract layer on the top of a middleware which hides communication and consistency management details from the game application. Hence the Synchronization Medium is a communication abstraction for game state synchronization. In this paper, we present the detailed architecture of the synchronization Medium with some improvements by adding more functionalities which help to achieve better synchronization between DIA users executing their applications on mobile terminals using a wireless network. \\

The paper is organized as follows. In section 2, we discuss the detailed architecture of the synchronization Medium 
ant its communication both with the game application and the underlying middleware. 
In section 3, we evaluate our synchronization Medium on three different  game applications. We discuss some related works in section 4 and conclude this paper and give future research directions in the last section.

\begin{figure}
\centering
\includegraphics[scale=0.7]{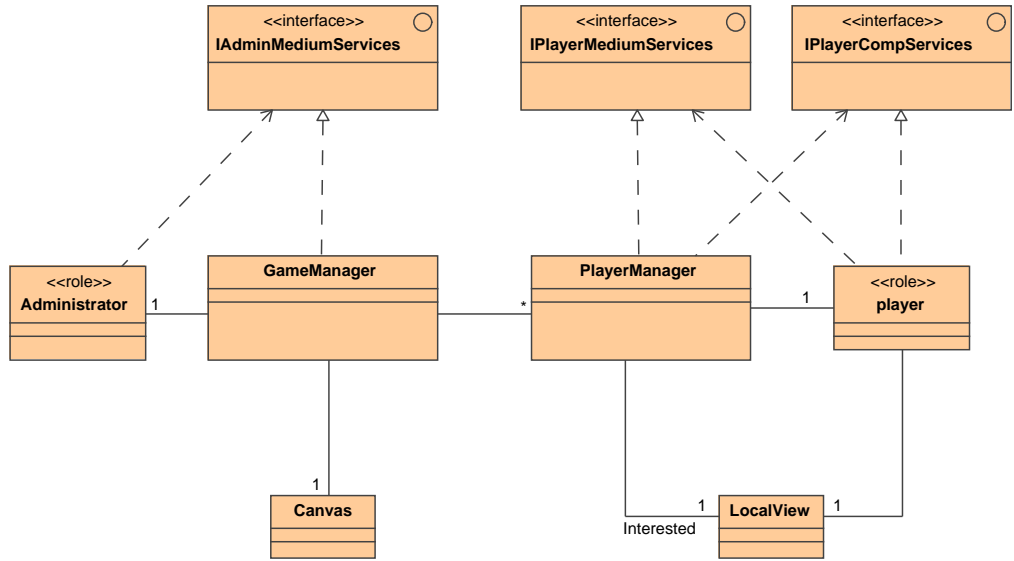}
\caption{Synchronization Medium}
\label{fig:Medium-layers}
\end{figure}

\section{Medium Detailed Architecture}
\begin{figure}
\centering
\includegraphics[scale=0.6]{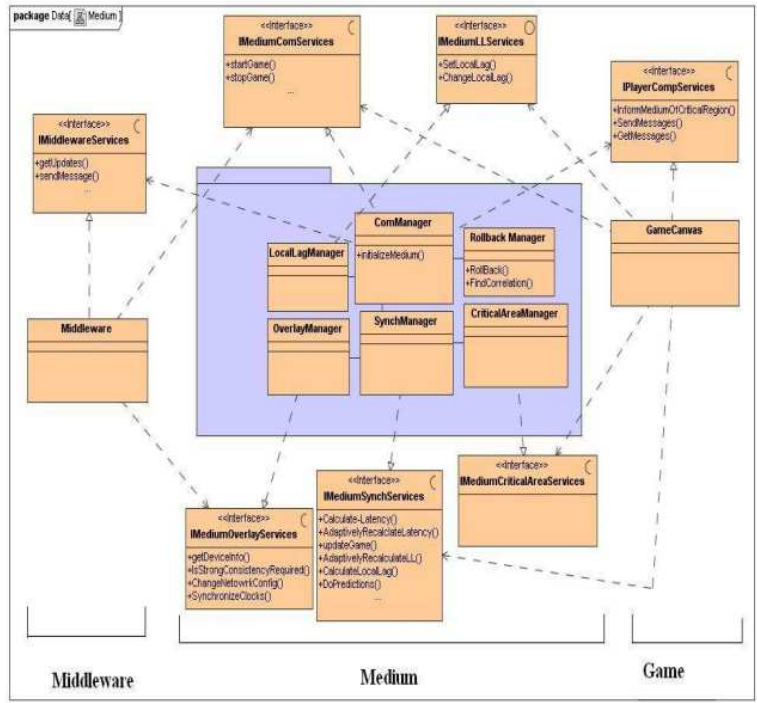}
\caption{Detailed Architecture of the Synchronization Medium}
\label{fig:Medium-detail}
\end{figure}
A Synchronization Medium is a composite component containing subcomponents. The Medium interacts with the game application by offering some services through its interfaces, and also requesting services from the game by using game services through the game interface. The part of the Medium residing on the client side is called a Player Manager. The set of all the player managers on all the clients playing a game constitutes the synchronization Medium. 

Each player manager is a composite component containing different managers. Figure \ref{fig:Medium-detail} shows the internal structure of the Medium and its interaction with the game application. In this section, we present the different managers of the synchronization Medium, how they interact with each other, and how they communicate with the game application and the underlying network infrastructure.
\subsection{Components of the Player Manager}
\paragraph{Critical Area Manager}
The Critical Area Manager (CAM) is responsible for the processing of all the information related to the critical region(s) in a game.  As stated previously, a critical region requires strict consistency for achieving better results. The CAM takes information from the game about the critical regions  and uses this information during the consistency management process. It offers its services through the \textit{IMediumCriticalAreaServices} interface. This interface is used by the game application to send information about the critical area to the CAM. This information can be the coordinates of the critical area such as the area around the base line in a tennis game or it can be the center of a circle of a certain radius around a player. In the latter case, the critical area is not fixed, but instead moves along with the player. In a critical region, the synchronization technique must be adapted to enforce strong consistency while loose consistency can be tolerated in some other \textit{less important} areas of the game environment. When a player enters a critical region, the CAM informs the Synchronization Manager (section 2.3) about the player's ID and its motion (speed, direction etc). It is then the responsibility of the Synchronization Manager to take the necessary actions so as to achieve strong consistency in this area. 

\paragraph{Communication Manager}
The Communication Manager is responsible for the communication
between the application logic on one client and the rest of the clients through an underlying middleware or network infrastructure. This includes, for example, the starting and stopping of the game application. It offers services to the underlying game middleware to communicate information with other remote game clients. This manager also adds a timestamp to each emitting message that is being sent to remote clients so that remote clients could process the arriving messages in temporal order as we discuss in section 2.2.

\paragraph{Synchronization Manager}
The Synchronization Manager (SM) is the part of the Medium which is responsible for the synchronization of game information and for state consistency management. 
A first hand separation of the synchronization concern is the calculation of network delays. We consider that this should be the responsibility of the synchronization Medium and not of the game itself. The SM is the Medium component that calculates the network latency. Based upon these latencies, the SM takes different actions such as prediction according to a certain Dead-reckoning algorithm. As we have shown in \cite{khan}, the network delays may change during the game play because of the network load, and hence the SM calculates network latency not only when the game starts, but also during the game play. This re-calculation of delays can be periodical or it can be reactive to certain network conditions such as a variation in the network load. If the SM is aware of the game map, such as the racing track provided by the game application, it can do the necessary prediction and then pass the predicted message to the game application. If the prediction is not possible in the synchronization Medium, because of insufficient game canvas information on the part of the Medium, then it is the responsibility of the game application to do predictions there. The game application uses the \textit{IPlayerMediumSynchServices} interface of the SM to benefit from the services offered by the Medium for the purpose of synchronization. 

\paragraph{Local Lag Manager}
The Local Lag Manager is responsible for taking the decision about the local lag corresponding to the delay produced artificially locally before the playout of commands \cite{Mauve}. As we have proposed an adaptable local lag mechanism in \cite{khan}, where the value of the local lag depends upon the network and game conditions, it is the responsibility of the Local Lag Manager to fix a value for the local lag and to decide when to change this value. As the local lag value depends upon the object about which the information is being received or displayed, it is the responsibility of the game application to inform the synchronization Medium about the initial value of the local lag for that object. Note that different objects in the game world can be assigned different local lag values according to their synchronization needs as we have discussed in \cite{khan}. The value of the local lag is also dependent on whether the object in question is in the critical region of the game or not. If the object is in the critical region, then decreasing the local lag value means using more real-time messages thus reducing the inconsistencies in the outcome of the game in that area.
 
\paragraph{Rollback Manager}
The Rollback Manager component has the responsibility for keeping the received messages in temporal order. It observes all the incoming messages, and if a message $m$ arrives late, then it informs the game through its \textit{IPlayerCompServices} interface to rollback all those messages whose timestamps are greater than that of $m$ i.e. those messages which were issued after $m$ by the emitting node but arrived earlier than $m$. This rollback can be combined with an approach such as optimistic correlation 
\cite{optimistic-correlation} to reduce the number of rollbacks on the receiving side. 
Note that the Rollback Manager does not offer any service to the game but only uses game services for rollbacks.

\paragraph{Overlay Manager}
The Overlay Manager (OM) is the component of the Medium responsible for taking those network related decisions which can affect the synchronization. Based on their position in the game, different players have different requirements. For example, in a football game, the consistency requirement for the players near the goal post (a critical region) is stronger than in other areas. So, there is a need to exchange information quickly between those players who are near the goal post as compared to the others. 
When exchanging messages during the game, the player sends information about that player's critical region. So this information should be sent only to the concerned players. This can be done directly (to avoid any delay) or through a server. In case the IP address of the remote terminal is known, a direct pair-to-pair connection is possible without passing through the server. As an example, in case of playing a soccer game on mobile phones, the Bluetooth technology could be used to communicate with nearby physical players if they are indeed nearby in the virtual world. This change of the network configuration for the sake of strong consistency during the game play is the responsibility of the OM. For the Medium to be able to do this, it should collect some information at the start of the game. This information includes remote terminal IP address, whether a terminal is equipped with a GPS, its memory size and processor power. In case the remote terminal and the local terminal are both equipped with GPSs, the OM can be used to synchronize their clocks so that the Synchronization Manager calculates the latency time between them by passing the timestamp in their message exchanges. Thus an OM is responsible for taking network related decisions during the game play, thereby improving the overall performance of the game. 

The OM can also decide to switch from one network to another during the game play. This decision can be based on different factors such as switching from a slower network to a faster one, or from a costlier one to a cheaper one, etc. 

\begin{figure}
\centering
\includegraphics[scale=0.45]{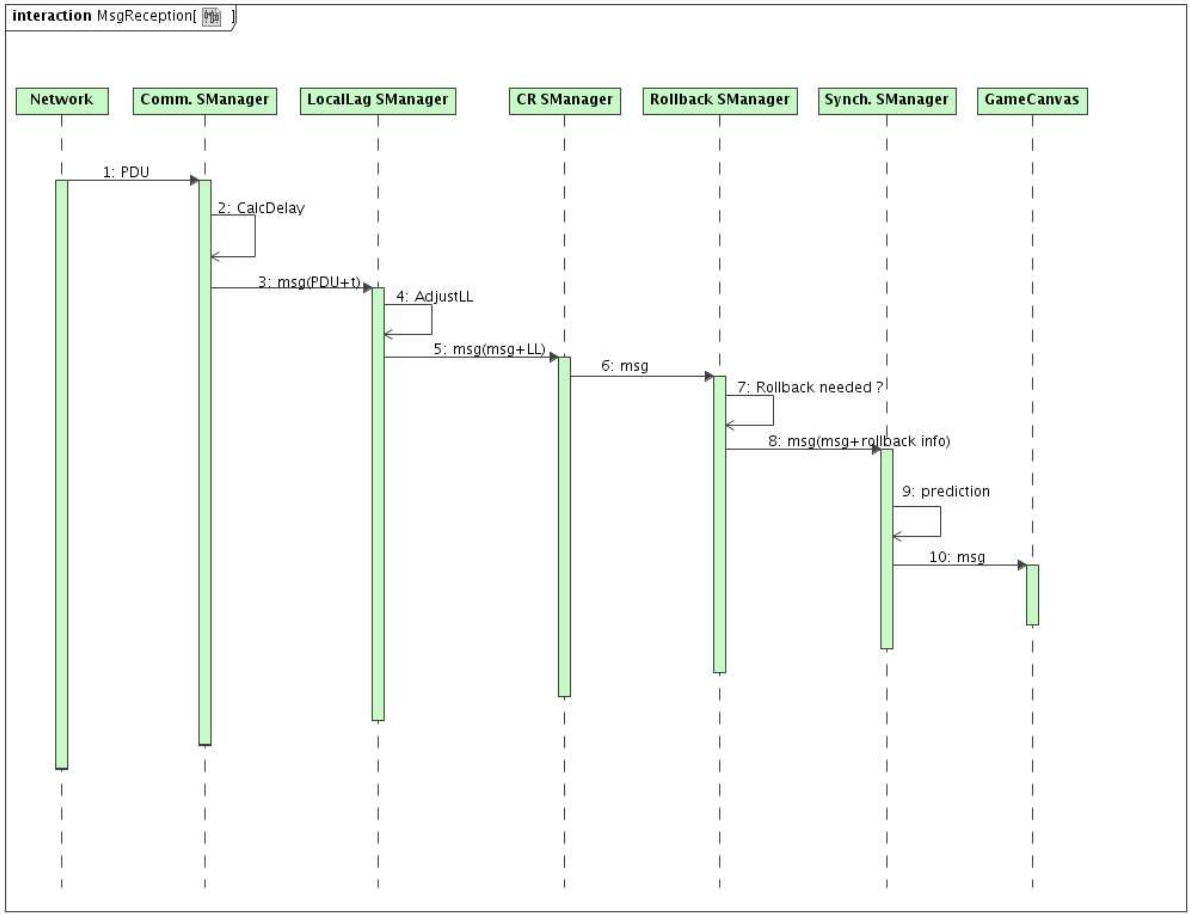}
\caption{Message Reception}
\label{fig:Medium-msg-reception}
\end{figure}

\subsection{Message Reception and transmission}
Figure \ref{fig:Medium-msg-reception} shows how remote messages are received by the Medium. All the messages received from a remote client are passed through the synchronization Medium. The Medium receives PDUs (protocol data units), containing information such as remote player's position, direction, speed and timestamp etc, from the underlying network or middleware. The message is received by the Communication Manager of the Medium. This Manager computes the delay either from the timestamp of the message in case the clocks are synchronized or through some other means at the start of the game or periodically during the game. This message along with the network latency is sent to the Local Lag Manager. The Local Lag Manager calculates the necessary lag to be introduced locally before sending the message. The Critical Region Manager decides whether a switch to strong consistency is necessary or not. The message is then passed to the RollBack Manager to decide whether a rollback is required depending upon whether the message was received on time or late. The message is then passed to the Synchronization Manager to apply a prediction algorithm before being passed to the game canvas.

Similiarly, all the messages sent by a game client to the other remote players are passed through the Synchronization Medium.  The message is first passed to the Synchronization Manager to apply local prediction. This local prediction is necessary for the client to decide when to send the next message. According the the dead-reckoning technique, the local client sends the message only when a local prediction error reaches a certain threshold. 
The message is then passed to the Critical Area Manager which detects if the player is in the critical area or not so that the Overlay Manager can decide whether a direct connection, through Bluetooth for instance if available, is necessary or not. The communication Manager then sends the message to the remote players over the network.  

\section{Evaluation} 
In this section, we present the evaluation work carried out on the synchronization Medium. We have followed the process below:

\begin{itemize}
\item implementation of a new game using the Medium
\item analysis of the reusability of the Medium for two different and pre-existing games
\item evaluation of the ease of development
\item and finally, performance evaluation.
\end{itemize}

The game logic relies on the underlying Medium layer for communication as well as for consistency maintenance. The Medium either uses the network infrastructure directly or through an underlying middleware to communicate with other remote players.

\subsection{Implementation of the Medium}
This section demonstrates the use of the synchronization Medium for the implementation of a car racing game that we developed. 
The software architecture we built is shown in the class diagram of Figure \ref{fig:MediumCar}.

\begin{figure}
\centering
\includegraphics[scale=0.35]{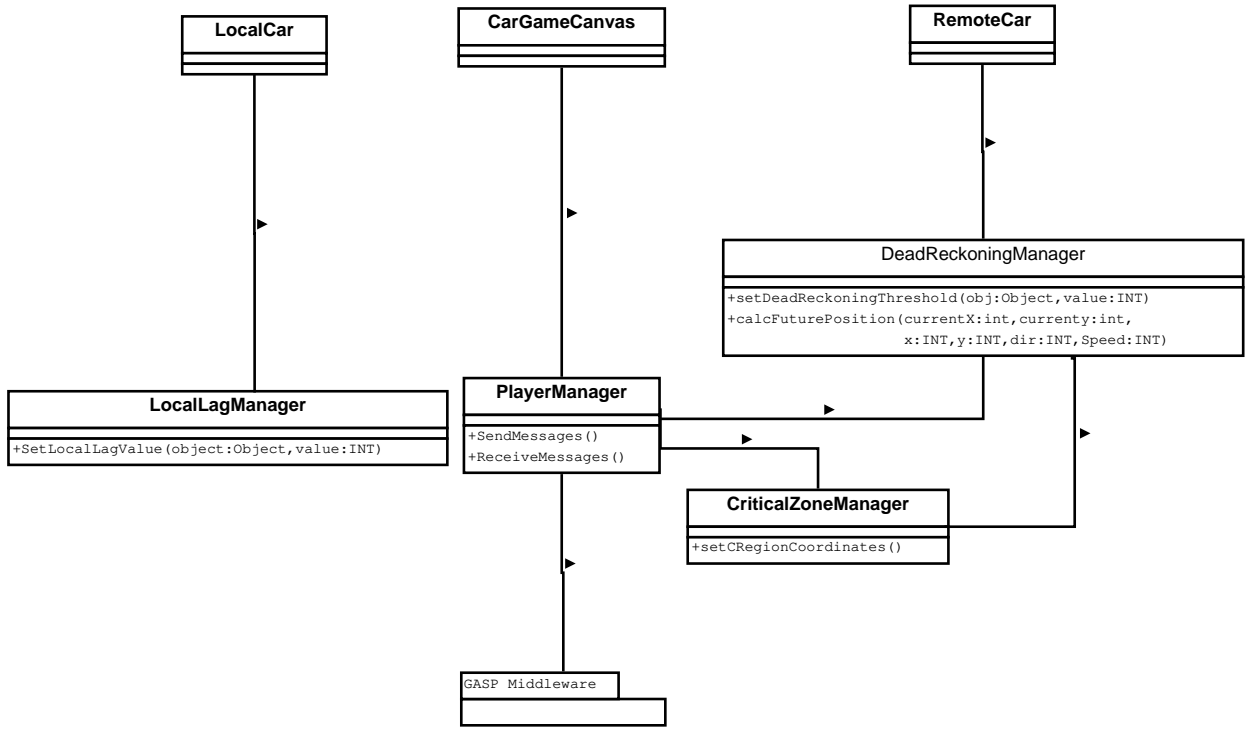}
\caption{A Medium as used by a car racing game}
\label{fig:MediumCar}
\end{figure}

The Medium is represented by four classes: The \textit{PlayerManager} class is the core Medium class and is responsible for the coordination amongst the other Medium classes and also with the game classes. It manages message reception from the remote players as well as message transmission to the network infrastructure. Note that for the interface with the underlying network infrastructure, we rely on the services of the GASP middleware \cite{gasp} dedicated to mobile games development. The \textit{DeadReckoningManager} class is responsible for applying the dead-reckoning algorithms and specifying their thresholds. Although dead-reckoning is dependent on the game player's direction, speed and the terrain in general, we can provide a synchronization Medium with a dead-reckoning Manager for all the games in a certain class of games. The \textit{CriticalZoneManager} class and the \textit{LocalLagManager} class are responsible for managing the critical regions and the local lag usage in the game. 

\subsection{Reusability of the Medium}
In this section, we argue that the architecture we propose is re-usable without modifying the underlying Medium's architecture.
We show the re-usability of the Medium by deploying two different multiplayer games, a Space War game and a multiplayer battle tank game, in addition to the car racing game, on the top of it without modifying the underlying Medium and middleware code. 

In figure \ref{fig:spaceWarOriginal}, we show the original SpaceWar game without any Medium. This is a multiplayer game that can be downloaded from \cite{powers}. It is developed in Jave ME \cite{J2MEGAMEPROGRAMMING}, the Sun Java platform for mobile devices. 

\begin{figure}
\centering
\includegraphics[scale=0.5]{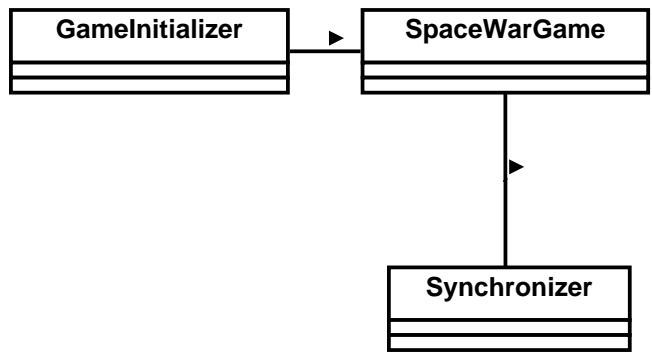}
\caption{SpaceWar game without using any Medium}
\label{fig:spaceWarOriginal}
\end{figure}

The original game has nine classes representing three different parts of the game: 1) the game initialization (GameInitializer) 2) the game application and display (SpaceWarGame) and 3) the communication part (Synchronizer). The GameInitializer class is responsible for initializing the game and performing the necessary administration tasks such as managing user accounts and passwords before starting the game. The communication part is represented by the \textit{Synchronizer} class responsible for communication as well as for some part of the consistency maintenance. \\

In Figure \ref{fig:spaceWarWithMedium}, we show how an existing Medium can be re-used by a game developer for the consistency maintenance of the SpaceWar game. Like in the previous section, the Medium is represented by a set of Managers. 
The game classes are now interacting with the Medium using the interfaces of the player manager. In the real implementation, this game has more than one class, but for the sake of simplicity, we only show the main canvas class which interacts with the Medium and manages the display on the client side. 
Moreover, when using the Medium, the GameInitializer (or Midlet in Java ME) is not required as it is the Medium which is responsible for initializing and starting the game by calling the midlet's 
\hbox{\textit{startApp()}} method \cite{J2MEGAMEPROGRAMMING}.  

\begin{figure}
\centering
\includegraphics[scale=0.33]{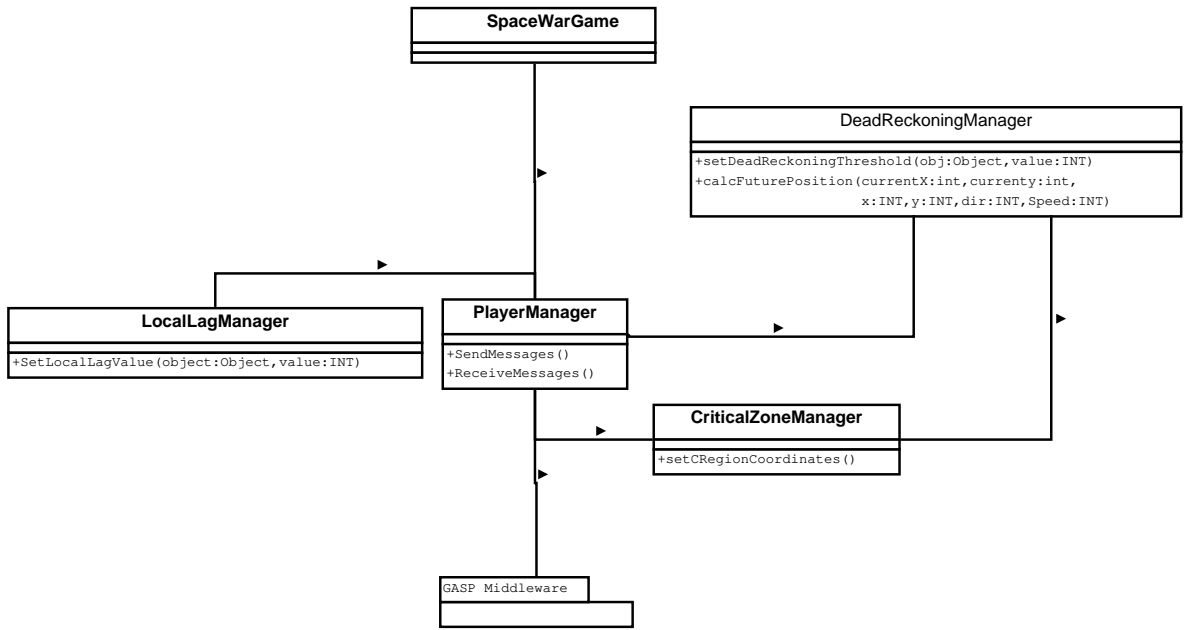}
\caption{SpaceWar game using the Medium}
\label{fig:spaceWarWithMedium}
\end{figure}

In the original game code, the code for the synchronization and consistency maintenance is embedded into different classes depending upon the developer's approach. For future enhancements or for using another consistency maintenance algorithm, we have to change much of the synchronization related code weaved into the game logic, which is costly and time consuming. On the other hand, when using the synchronization Medium, we write the code in a class which is not part of the game logic and which can be easily re-used. The communication and synchronisation issues become of the responsibility of the Medium and of the underlying middleware. Hence, the developer is relieved from developing this part of the game and can then concentrate fully on the game itself.

With the use of the Medium (figure \ref{fig:spaceWarWithMedium}), whenever a message is received by the underlying infrastructure, it is accepted by the Medium's \textit{PlayerManager} class. This class calculates the time this message has taken from its delivery at the sender's terminal till its reception by this class, supposing that the clocks of all the players are synchronized using some clocks method such as Network Time Protocol (NTP) \cite{Mills89}, GPS \cite{GpsSynchronization} or any other method \cite{OtherClockSynchronization}. The Medium buffers the message for a specific time determined by the local lag threshold value. This threshold value is provided by the developer of the game by implementing the  \textit{setLocalLagValue()} abstract method of the \textit{LocalLagManager} class of the synchronization Medium. The message is then sent to the game canvas class of the game. The game canvas calls the method \textit{calcFuturePosition()} of the \textit{DeadReckoningManager} class of the Medium to apply dead-reckoning algorithm for the calculation of the correction future position of this update message. This is necessary, because the message belongs to the past and now the sender of the message has moved to a new place.

The \textit{Critical Area Manager} is responsible for defining the critical regions in the game. The critical regions are defined for each object and for the game terrain. These regions are defined by the \hbox{\textit{SetCRegionCoordinates()}} method of the \textit{Critical Region Manager}. In our first implementation of the synchronization Medium with a car racing game, we have a single critical region and that is an area of the track just before the arrival line. When a car enters that region, the game canvas signals it to the player manager which then changes the values of dead-reckoning threshold and local lag accordingly to achieve a strong consistency. Through the interface provided by the Medium for a specific genre of games, the game developer must provide the coordinates of the critical regions to the Medium. 
 
The \textit{Local Lag Manager} is responsible for setting the local lag value for each class of objects and changing it dynamically according to the objects' pace and position in the game. In our implementation of the car racing game with just two cars, we kept this value equal to the maximum network delay between the two players. It was 500 ms and 1000 ms in two different experiments. This value is set by the game developer by implementing the \hbox{\textit{setLocalLagValue()}} method of the \textit{Local Lag Manager}. As this is an abstract method, the developer has to implement it. \\

In order to confirm the reusability of our approach, we proposed a project to a student not familiar with synchronization or Mediums. The student transformed a single player battle tank game into a multiplayer game reusing the synchronization Medium architecture \cite{jml-ref}. A class diagram of the multiplayer version of the battle tank game using the synchronization Medium is shown in figure \ref{fig:tankmultiplayer}. In the game, each player owns a tank and tries to explode the tank of an opponent by targeting it with tank bullets. The main game logic classes of the battle tank game are $Tank$, $RemoteTank$, $HeroTank$, $Bullet$ and $Explosion$. They use different managers of the synchronization Medium to communicate and synchronize with remote players.  
This project gives us confidence in the reusability and modularity of our approach. The deployment of three different multiplayer games (car race, space war and battle tank) on the top of the synchronization Medium confirms its reusability.

\begin{figure}
\centering
\includegraphics[scale=0.55]{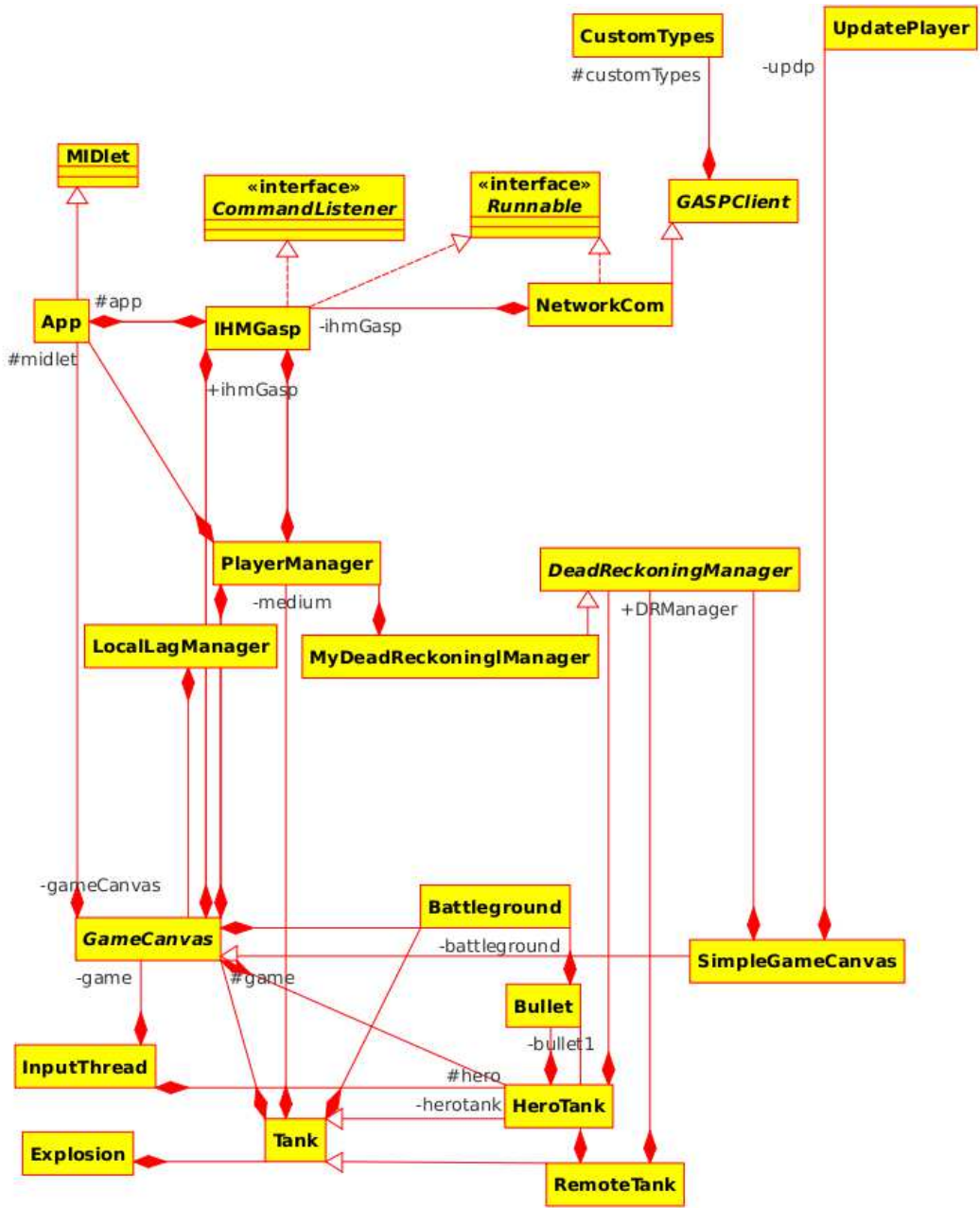}
\caption{Class diagram of a multiplayer tank game using the Medium}
\label{fig:tankmultiplayer}
\end{figure}

\subsection{Comparison of development efforts}
In this section, we compare quantitatively the efforts spent for the implementation of a game with and without Medium. We argue that the proposed architecture lessens the development efforts in terms of the number of classes and lines of code. \\
Table \ref{tb:comparison} shows a comparison of the development efforts when implementing the game on the top of the Synchronization Medium. As can be seen, the number of core game logic classes when using Synchronization Medium is less than when the game is developed without the Synchronization Medium (case of the Space War game ). The Medium classes (related to synchronization and communication) are reused by the game and the programmer is relieved of writing the lines of code responsible for consistency maintenance. 

\renewcommand{\arraystretch}{1.4}

\begin{table}
\caption{Comparison of development efforts with and without a Synchronization Medium}
\label{tb:comparison}
\begin{center}
\begin{tabular}{|c|c|p{3cm}|p{3cm}|}

 \hline
     Game & Initial Status & Game Logic Classes Using Medium & Reused Code (Medium Classes)\\  \hline
    Car Race & None - Developed game & 6 &  4 \\ \hline
    Space War & Available with 9 classes & 7 & 4 \\  \hline
    Battle Tank & Available as MonoPlayer game & 8 & 4 \\ \hline
\end{tabular}
\end{center}
\end{table}

\subsection{Performance Evaluation}

In this section, we show that the insertion of a new consistency layer below the game logic classes does not introduce any untoward inconsistency, for example, due to an increase of the message processing time.
Table \ref{tb:comparison2} shows a comparison of the delay of message processing with and without a Synchronization Medium. For an average network delay of 500 ms to 1000 ms, the average time for message processing is 0-5 ms when using the Medium as compared to 0-2 ms when the game is not using the Synchronization Medium. It means that, for an average delay of 500 ms, the additional processing time is only one percent at most as compared to a non-Medium game implementation.
This is not a significant delay as compared to the advantages in terms of game architecture and development time.

\begin{table}
\caption{Qualitative comparison with and without a Synchronization Medium}
\label{tb:comparison2}
\begin{center}
\begin{tabular}{|c|p{5cm}|p{5cm}|}
 \hline
     Network Delay & Stand-alone Game message processing time & Processing time for message when game is using Synchronization Medium\\  \hline
    500ms & 0-2ms & 0-5ms \\ \hline
    1000ms & 0-2ms & 0-5ms \\ \hline
    
\end{tabular}
\end{center}
\end{table}

In figure \ref{fig:tankbullet}, we compare the positions of the bullets of a battle tank (as discussed in the previous section) locally and when displayed at a remote player using a certain local lag value.
The average difference, in terms of display time, between the local bullet positions and when displayed on the remote site is 21,5ms which is well in the acceptable range of human perception. Hence, the insertion of a Medium layer into the game logic code does not introduce performance degradation.
\begin{figure}
\centering
\includegraphics[scale=0.7]{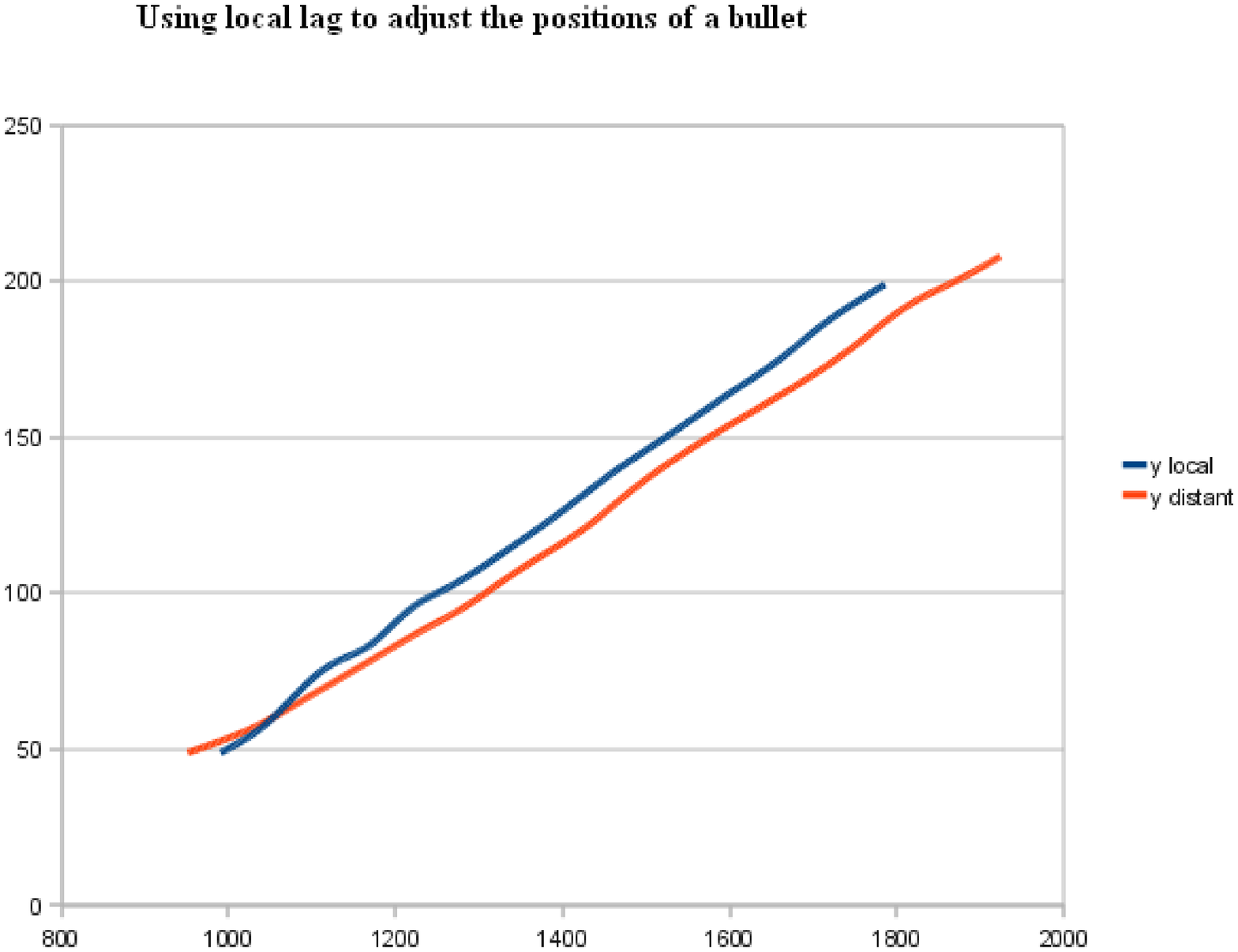}
\caption{Position of a tank bullet at local and remote tank}
\label{fig:tankbullet}
\end{figure}

\section{Related Works}
\cite{fletcher06} proposes a Concurrency Control and
Consistency Maintenance (CCCM) component to handle consistency issues
separately from the game logic. The CCCM component implementing the consistency management
algorithms resides between the game logic and the
game data. We take a step further by decoupling the synchronization
issues completely from the game logic and data and injecting it into a
communication component which handles the consistency management and
returns the results to the players. Another important difference
with our approach is that CCCM component is
for client server architectures, while the synchronization Medium is
not limited to a centralized architecture. From the same abstract
specification we can implement a synchronization Medium for
Peer-to-Peer or PP-CA (Peer-to-Peer with central arbiter)\cite{pp-ca}
architecture during the design process.

The Mammoth research framework has been proposed recently \cite{KVKDH09} 
to evaluate different design assumptions for massively multiplayer
games. This framework is based on a collection of cooperating components
providing modularity and flexibility like our proposal. The difference
lies in the way consistency is handled. Consistency in Mammoth  
is guaranteed by designating one of the copies of the duplicated 
objects as being the duplication master. This is 
sufficient for multiplayer games relying on wired networks with limited 
latency. In order to be able to deal with higher latencies such as in wireless networks,
we consider
consistency management as a first-order design issue and we propose
a synchronization Medium where various algorithms can be plugged in.

\section{Conclusions and perspectives}
 
In this paper, we presented a composite component for communication and 
consistency maintenance in mobile multi-user virtual environments, such as 
multiplayer games. Our proposed architecture is reusable as it provides generic 
interfaces to game developers. We demonstrated its reusability using three 
different multiplayer games. We also evaluated the overhead introduced by the 
Medium and showed that it remains acceptable.

In the future, we plan to implement different synchronization Mediums for 
different architectures (e.g. Mirrored Server or peer-to-peer
architectures) to be used by the same game application. In this way, we would be 
able to show that the game application can be run on different platforms 
 by using off-the-shelf Mediums, without resorting to change the game code.

\bibliographystyle{abbrv}
\bibliography{paper}
\end{document}